\documentclass[12pt]{article}  
\newcommand {\bea}{\vspace{-.00in}\begin{eqnarray}}   
\newcommand {\eea}{\vspace{-.00in}\end{eqnarray}}
\newcommand {\beaa}{\vspace{-.00in}\begin{eqnarray*}}   
\newcommand {\eeaa}{\vspace{-.00in}\end{eqnarray*}}
\newcommand {\be}{\vspace{-.00in}\begin{equation}}   
\newcommand {\ee}{\vspace{-.00in}\end{equation}}

\newcommand {\del} {\delta}

\newcommand  {\lam} {\lambda}
\newcommand  {\pr} {\prime}

\newcommand  {\Lam} {\Lambda}

\newcommand   {\pp}   {({\bf p}-{{\bf p}^\pr})^2}
\newcommand   {\easy}  {\del M_{_{1}}^2}
\newcommand    {\hard} {\del M_{_{2}}^2}

\begin{document}
\thispagestyle{empty}
\centerline{\large {\bf  Singlet-Triplet Splitting of Positronium in
 Light-Front QED 
}}
\vskip.2in
\centerline{Billy D. Jones}
\centerline{  {\em Department of Physics, The Ohio State University, Columbus OH 
43210}}
\centerline{(March 14, 1997)}
\vskip.2in
\centerline{\large Abstract}
\vskip.15in
\noindent
 We study the QED bound-state problem  in
 a  light-front
  hamiltonian approach. It is important to establish the equivalence (or
  not)
  of equal-time and light-front approaches
  in the well-understood arena of
  Quantum Electrodynamics. Along these lines, the singlet-triplet ground state
  spin splitting in positronium is calculated. The
  well-known result, $\frac{7}{6} \alpha^2 Ryd$, is obtained analytically, which
  establishes the equivalence between the equal-time and light-front approaches 
(at least to
  this order).
   The true equivalence of the two approaches 
  can only be established after higher-order calculations. It was 
  previously shown that this light-front
  result could be obtained analytically \cite{jones}, but
  a simpler method is presented in this paper.
 \vskip.2in
\noindent
Talk presented at {\em Orbis Scientiae 1997}, Miami Beach, Florida, January 
23--26, 1997.
To appear in the proceedings.
\newpage
\centerline{{\bf I. INTRODUCTION}}
\vskip.2in

A calculation of the singlet-triplet ground state spin splitting in positronium
is rather trivial from the viewpoint of a Coulomb gauge equal-time calculation
(see for example \S83--84 of \cite{qed}). This is not the case in a 
light-cone gauge
light-front calculation (see for example \cite{kaluzapirner}).
 We will briefly
outline the derivation of the effective Hamiltonian, and then proceed with
an analytic calculation of the singlet-triplet splitting.

\vskip.2in
\centerline{{\bf II. EFFECTIVE HAMILTONIAN AND ITS ZEROTH ORDER SPECTRUM}}
\vskip.2in

A brief description of the approach will be given, and the resulting
effective Hamiltonian which will be studied in this paper will be written.
For details of the derivation of the effective Hamiltonian,
and for the original
references, see \cite{jones}. 

The starting point is the canonical QED Hamiltonian in the light-cone gauge.
Then a regulator is introduced, $\Lam$, which removes high energy exchanges
from the theory. Proceeding, a unitary transformation is defined that acts on
the regulated canonical Hamiltonian and produces an effective Hamiltonian at
a lower energy scale, $\lam$. The transformation is unitary, so the spectrum of
the regulated canonical Hamiltonian and the effective Hamiltonian are equivalent
(of course approximations can invalidate this conclusion). The $\Lam 
\longrightarrow \infty$
limit is studied, and the regulated canonical Hamiltonian is adjusted so that 
this
limit can be taken. So far this is nothing but
the old story of renormalization, but the procedure is far from trivial
in a light-front approach since longitudinal locality is lost.\footnote{
On a positive note, recall that there is an exact scale invariance of the theory
under a longitudinal scaling; thus no nonperturbative longitudinal scale can 
arise
through the process of renormalization.} Unfortunately a complete
story of how the renormalization works out is not available, the study is
very much a work in progress.
 However, to obtain the results of this paper consistently,
only the one loop electron self-energy renormalization needs to be performed. 
This
will not be shown explicitly in this paper, but is in \cite{jones}. The 
resulting
effective Hamiltonian satisfies the  Schr{\"o}dinger equation written below, 
which is
conveniently written in the notation of one body Quantum Mechanics. Note, to
obtain this form of the second order effective Hamiltonian, we needed to place
the scale in the following window, $m \alpha^2 \ll \lam \ll m \alpha$. This 
lower
bound is the nonperturbative energy scale of interest; if $\lam$ is lowered 
below this bound,
the Coulomb interaction does not arise from the second order effective 
interactions alone. This
upper bound is the dominant energy of emitted and absorbed photons;
placing $\lam$ below this bound allows the leading
order results
to be obtained in the valence sector alone. Given this, the second order 
effective
Hamiltonian satisfies the following Schr{\"o}dinger equation

\bea
\left(\hat{{\cal H}}_o+\hat{{\cal V}}
\right) |\Phi_N\rangle&=& M_N^2 |\Phi_N\rangle
\;,
\label{eq:sadd}
\eea  
where $M_N$ is the mass of the state and
\bea
&\bullet& \langle \Phi_N |\Phi_{N^\pr}\rangle = \del_{N N^\pr}\\
&\bullet& 1 = \sum_{s_1 s_2} \int d^3p \;|{\bf p} s_1 s_2 \rangle \langle {\bf 
p} s_1 s_2 | =
\sum_{s_1 s_2} \int d^3x\; |{\bf x} s_1 s_2 \rangle \langle {\bf x} s_1 s_2 |=
\sum_{N} |\Phi_N \rangle \langle \Phi_N|\\
&\bullet& \langle {\bf p}^\pr s_3 s_4 | \hat{{\cal V}} 
| {\bf p} s_1 s_2 \rangle = {\cal V}({\bf p}^\pr s_3 s_4;
 {\bf p} s_1 s_2)\;\;\\
&\bullet&\langle {\bf p}^\pr s_3 s_4 | \hat{{\cal H}}_o | {\bf p} s_1 s_2 
\rangle =
4 (m^2+{\bf p}^2)\del^3(p-p^\pr) \del_{s_1 s_3} \del_{s_2 s_4} -(4 m) 
\frac{\alpha}{2 \pi^2}
\frac{ \del_{s_1s_3}\del_{s_2s_4}
  }{({\bf p}-{\bf p}^\pr)^2}\\
&\bullet& M_N^2= (2 m+B_N)^2
\;\label{eq:101}.
\eea
$m$ is the electron mass,
$-B_N$ is the binding energy, and
$N$ labels all the quantum numbers of the state. For notational purposes note 
that
we label the final relative three-momentum with a prime, and that the initial 
and final
electrons
are labeled by ``1"  and ``3" respectively, and the initial and final positrons 
are labeled by
``2" and ``4" respectively. Before proceeding to write $\hat{{\cal V}}$, it is 
convenient
to discuss the spectrum of $\hat{{\cal H}}_o$. 

In zeroth order $\hat{{\cal V}}$ is neglected and Eq.(\ref{eq:sadd}) becomes
\bea
&&\hat{{\cal H}}_o
 |\phi_N\rangle = {\cal M}_N^2 |\phi_N\rangle = \left(
 4 m^2 + 4 m {\cal B}_N\right) |\phi_N\rangle \label{eq:zero}~.
 \eea
 This last equality defines our zeroth order binding energy, $-{\cal B}_N$.
 Projecting this eigenvalue equation into momentum space gives
 \bea
 \left(-{\cal B}_N+\frac{{{\bf p}^\pr}^2}{m}
 \right) \phi_N({\bf p}^\pr  s_3 s_4)&=&
 \frac{\alpha}{2 \pi^2}\int\frac{ d^3 p }{({\bf p}-{\bf p}^\pr)^2}
  \phi_N({\bf p} s_3 s_4)
 \;,\label{eq:C}
\eea
the familiar non-relativistic Schr{\"o}dinger equation for positronium.

After a simplification detailed in the next Section and mentioned in the
Abstract, like the Coulomb gauge equal-time calculation, to obtain
the ground state singlet-triplet 
splitting, only the wave function at the origin is required, which we
thus record:
\bea
\left(\phi_N\left({\bf x}=0\right)\right)^2&=& \frac{1}{(2 \pi)^3}
\left(\int d^3 p ~\phi_N\left({\bf p}\right)\right)^2=\frac{1}{\pi} \left(
\frac{m \alpha}{2 n}\right)^3 \del_{l,0}
\;.\label{eq:hy}
\eea
$n$ is the principal quantum number, and $l$ is the angular momentum quantum 
number.

Now we proceed to write $\hat{{\cal V}}$. Note that $\hat{{\cal V}}$ is not
diagonal in momentum space, so if we define
\bea
\hat{{\cal V}}&=&\hat{{\cal V}}^{^{(0)}}+\hat{{\cal V}}^{^{(1)}} +\hat{{\cal 
V}}^{^{(2)}}
+ \cdots\;,\label{eq:yes}
\eea
where the superscript implies the $\alpha$-scaling of a matrix element of
the operator in momentum space, then in
first order bound-state perturbation theory these operators contribute
\bea 
\langle \phi_N|\hat{{\cal V}}^{^{(S)}}|\phi_{N^\pr} \rangle 
\sim m^2 \alpha^{3 + S}
\;.\label{eq:80}
\eea
Note that Eq.~(\ref{eq:yes}) starts at $S=0$; this is a result of the
derivation of the effective Hamiltonian. Interestingly, note that in a
Coulomb gauge equal-time calculation, this series starts at $S=1$.
In summary, to be consistent to order $\alpha^4$, we need to look at all  matrix 
 elements 
 ${\cal V}^{^{(S)}}({\bf p}^\pr s_3 s_4;{\bf p} s_1 s_2)$ 
  with $S \leq 1$.\footnote{
For example,
 $ \frac{m e^2 {\bf p}}
 {({\bf p}-{\bf p}^\pr)^2} 
 \sim\frac{\alpha^2}{\alpha^2} 
 \Longrightarrow S = 0\;$.}
 
 Before proceeding to write out these expressions for $\hat{{\cal V}}^{^{(S)}}$,
 note that we only calculate spin splittings, so constants along the diagonal
 in spin space were neglected. Given this, to get the spin splittings correct
 to order $\alpha^4$ we need to consider
 \bea
 {\cal V}^{^{(0)}}({\bf p}^\pr s_3 s_4;
 {\bf p} s_1 s_2)&=& \frac{- c_{_{ex}} e^2}{4  \pi^3 ({\bf p}-{\bf p}^\pr)^2}
  { v}^{(0)}({\bf p}^\pr s_3 s_4;
 {\bf p} s_1 s_2)
 \;,\label{eq:rzero}
 \eea
 where
 \bea
 { v}^{(0)}({\bf p}^\pr s_3 s_4;
 {\bf p} s_1 s_2)&=&\left(\del_{s_1 {\overline s}_3} \del_{s_2 s_4} f_1({\bf 
p}^\pr s_3 s_4;
 {\bf p} s_1 s_2)+
 \del_{s_1 s_3} \del_{s_2 {\overline s}_4} f_2({\bf p}^\pr s_3 s_4;
 {\bf p} s_1 s_2)\right)\label{eq:132}~,\\
 f_1({\bf p}^\pr s_3 s_4;
 {\bf p} s_1 s_2) &=& 
 s_1 (p_y-p_y^\pr)-i (p_x-p_x^\pr) ~,\\
 f_2({\bf p}^\pr s_3 s_4;
 {\bf p} s_1 s_2) &=& s_4 (p_y-p_y^\pr)+i (p_x-p_x^\pr) 
 \;.
 \eea
  $s_i/2$ is the spin quantum number of fermion ``i"; $s_i = \pm 1~(i=1,2,3,4)$ 
only;
  ${\overline{s}}_i=-s_i$.
  The only other interaction that needs to be considered is
 \bea
 {\cal V}^{^{(1)}}({\bf p}^\pr s_3 s_4;
 {\bf p} s_1 s_2)&=& \frac{e^2}{4 m \pi^3}\left(c_{_{an}} \del_{s_1 s_2} 
\del_{s_1 s_4}
  \del_{s_3 s_4}+ c_{_{ex}} \del_{s_2 {\overline s}_4} \del_{s_2 {\overline 
s}_1}
  \del_{s_3 {\overline s}_1}+\right.\nonumber\\
 &&~~~~+ \left.\left(c_{_{an}} \frac{1}{2} - c_{_{ex}}
 \frac{(p_\perp-p_\perp^\pr)^2}{({\bf p}-{\bf p}^\pr)^2}\right)
  \del_{s_1 {\overline s}_2} \del_{s_3 {\overline s}_4}\right)\;.
 \eea
 The constants $c_{_{ex}}$ and $c_{_{an}}$
  were introduced only to distinguish the terms that arise from the `exchange' 
and 
 `annihilation' channels respectively; $c_{_{ex}} = c_{_{an}} = 1$.
 
\vskip.2in
\noindent
\centerline{{\bf III. SINGLET-TRIPLET SPLITTING}}
\vskip.2in

Now we will calculate the ground state singlet-triplet splitting to order
$\alpha^4$ using bound-state perturbation theory in $\hat{{\cal V}}$. 
Perhaps the most straightforward approach is to just get busy and calculate,
since the non-relativistic Coulomb spectrum is so well known. This is
exactly what is done in \cite{jones}; however, as can be seen by the complexity
of Appendix~C in that paper, the calculation is complicated and at the
level of a ``Lamb shift calculation." We will now present a simpler method
to calculate this shift.\footnote{The idea behind
  this simpler method originated
  with Brisudov\'{a} and Perry \cite{m1}.} This simpler method uses a unitary 
transformation
to ``remove" $\hat{{\cal V}}^{^{(0)}}$ much in the spirit of Schwinger's early 
QED
calculations \cite{schwinger}. This simpler method now follows.

First, set up a general unitary transformation with hermitian generator 
$\hat{Q}$:
\bea
  \hat{H}&=& \hat{{\cal H}}_o+\hat{{\cal V}}^{^{(0)}}+
 \hat{{\cal V}}^{^{(1)}}+\hat{{\cal V}}^{^{(2)}}+\cdots~,\\
 \hat{H}^\pr&=& e^{i\hat{Q}} \hat{H} e^{-i\hat{Q}}\nonumber\\
 &=&\hat{H}+i\left[\hat{Q}, \hat{H} 
\right]+\frac{i^2}{2!}\left[\hat{Q},\left[\hat{Q},\hat{H}\right]
 \right]+\cdots \;.\label{eq:hprime}
 \eea
 Now define $\hat{Q}$ by requiring its commutator with $\hat{{\cal H}}_o$ to 
cancel 
 $\hat{{\cal V}}^{^{(0)}}$:
 \bea
 \hat{{\cal V}}^{^{(0)}}+ i \left[
 \hat{Q},\hat{{\cal H}}_o
 \right]&=&0
 \;.
 \label{eq:elimination}
 \eea
 Putting this into Eq.~(\ref{eq:hprime}) gives
  \bea
  \hat{H}^\pr&=& \hat{{\cal H}}_o+\left(1-\frac{1}{2!}\right)
   \left[ i \hat{Q},\hat{{\cal V}}^{^{(0)}}\right]+
  e^{i\hat{Q}} \left(\hat{{\cal V}}^{^{(1)}}+\hat{{\cal 
V}}^{^{(2)}}+\cdots\right) e^{-i\hat{Q}}
  \nonumber\\
 && ~+\left(\left(\frac{1}{2!}-\frac{1}{3!}\right)\left[i \hat{Q},
 \left[i \hat{Q},\hat{{\cal V}}^{^{(0)}}\right]
 \right]+
 \left(\frac{1}{3!}-\frac{1}{4!}\right) 
 \left[ i \hat{Q},\left[ i \hat {Q},\left[ i \hat{Q}, \hat{{\cal 
V}}^{^{(0)}}\right]
 \right]\right]+\cdots\right)
  \;.
  \eea
  Note that $\hat{H}$ and $\hat{H}^\pr$  have equivalent lowest order spectrums
  given by
  $\hat{{\cal H}}_o$;
  this can be seen easily by looking at matrix elements of the equations in 
Coulomb states,
  that is in states of $\hat{{\cal H}}_o$. To summarize,  we must diagonalize 
the following
  interaction in spin space to obtain the  order $\alpha^4$ 
  ground state singlet-triplet splitting in positronium:
  \bea
 \del^{^{(1)}}M^2(s_3,s_4;s_1,s_2)=\langle\phi_{1,0,0,s_3,s_4}|
 \hat{{\cal V}}^{^{(1)}}+\frac{1}{2} \left[ i \hat{Q},\hat{{\cal 
V}}^{^{(0)}}\right]
 |\phi_{1,0,0,s_1,s_2}\rangle
 \;,\label{eq:FFF}
 \eea
 where $\hat{Q}$ is a solution to Eq.~(\ref{eq:elimination}).
 The superscript on $\del^{^{(1)}}M^2$ signifies that it is a {\em first} order
 bound-state perturbation theory shift.
 The quantum numbers are 
 $N=(n,l,m_l,s_e,s_{\overline{e}})\longrightarrow (1,0,0,s_e,s_{\overline{e}})$
 for the ground state.
 
 In what follows we will solve Eq.~(\ref{eq:elimination}) for $\hat{Q}$ in the 
free
 basis in momentum space,\footnote{This is the trick, to solve for $\hat{Q}$ in 
the
 free basis; if $\hat{Q}$ is solved for in the Coulomb basis the calculation 
follows
 the one carried out in \cite{jones}.} and  then calculate the shift defined by 
 Eq.~(\ref{eq:FFF}). 
 
 From the form of $\hat{{\cal V}}^{^{(0)}}$ and $\hat{{\cal H}}_o$ we 
 see that $\hat{Q}$ has the following general form
 \bea
 \langle {\bf p}^\pr s_3 s_4|i \hat{Q}|{\bf p} s_1 s_2 \rangle&=&
 \del^3(p-p^\pr) \langle {\bf p}^\pr s_3 s_4|i \hat{R}|{\bf p} s_1 s_2 
\rangle\;,
 \eea
 where from Eq.~(\ref{eq:elimination}), 
 $\hat{R}$ satisfies
 \bea
 \frac{{ v}^{(0)}({\bf p}^\pr s_3 s_4;
 {\bf p} s_1 s_2)}{2 m}&=&
 \langle {\bf p} s_3 s_4|i \hat{R}|{\bf p} s_1 s_2 \rangle-
 \langle {\bf p}^\pr s_3 s_4|i \hat{R}|{\bf p}^\pr s_1 s_2 \rangle\;.
 \label{eq:hu}
 \eea
 Recall Eq.~(\ref{eq:132}) for the form of ${ v}^{(0)}$. Thus, the general
 form of $\hat{R}$ is
 \bea
 \langle {\bf p} s_3 s_4|i \hat{R}|{\bf p} s_1 s_2 \rangle&=&
\frac{\del_{s_1 {\overline s}_3} \del_{s_2 s_4}}{2m}\left(s_1 p_y-i p_x\right)+ 
 \frac{\del_{s_1 s_3} \del_{s_2 {\overline s}_4}}{2m}\left(s_4 p_y+i 
p_x\right)\;.\eea
 
 Since $\hat{Q}$ is diagonal in momentum space it is a simple matter to 
calculate
 the contributions from Eq.~(\ref{eq:FFF}). Define
 \bea
 \del M_{_{1}}^2&=&\langle\phi_{1,0,0,s_3,s_4}|
 \hat{{\cal V}}^{^{(1)}}
 |\phi_{1,0,0,s_1,s_2}\rangle~,\label{eq:178}\\
 \del M_{_{2}}^2&=&\langle\phi_{1,0,0,s_3,s_4}|
 \frac{1}{2} \left[ i \hat{Q},\hat{{\cal V}}^{^{(0)}}\right]
 |\phi_{1,0,0,s_1,s_2}\rangle
 \;.\label{eq:179}
 \eea
  
  First, $\easy$:  
 \bea
 \del M_{_{1}}^2&=&  \int d^3 p d^3 p^\pr \langle \phi_{100}|{\bf 
p}^\pr\rangle\langle{\bf
 p}|\phi_{100}\rangle 
  {\cal V}^{^{(1)}}({\bf p}^\pr s_3 s_4;
 {\bf p} s_1 s_2)
 \;.
 \eea
 Using the rotational symmetry of the integrand, we can replace
 \bea
 \frac{(p_\perp-p_\perp^\pr)^2}{\pp}&\longrightarrow&\frac{\frac{2}{3}\left(
 (p_x-p_x^\pr)^2+(p_y-p_y^\pr)^2+(p_z-p_z^\pr)^2\right)}{\pp}
 =\frac{2}{3}
 \;.
 \eea 
 After this,  the remaining integrals are trivial (recall Eq.~(\ref{eq:hy})) and 
we have
 \bea
 \frac{\easy}{2 m^2 \alpha^4}&=& \frac{1}{2} \del_{s_1 s_2} \del_{s_1 s_4} 
\del_{s_3 s_4} 
 -\frac{1}{12} \del_{s_1 {\overline s}_2} \del_{s_3 {\overline s}_4}+\frac{1}{2} 
 \del_{s_2 {\overline s}_4} \del_{s_1 {\overline s}_2} \del_{s_1
  {\overline s}_3}
  \;.
 \eea
 
 Next,  $\hard$: 
 \bea
 \hard&=&\langle\phi_{1,0,0,s_3,s_4}|
 \frac{1}{2} \left[ i \hat{Q},\hat{{\cal V}}^{^{(0)}}\right]
 |\phi_{1,0,0,s_1,s_2}\rangle\\
 &=&\frac{1}{2}\sum_{s_e s_{\overline{e}}}
 \int d^3 p d^3 p^\pr \langle \phi_{100}|{\bf p}^\pr\rangle\langle{\bf 
p}|\phi_{100}\rangle
 \left(\langle {\bf p}^\pr s_3 s_4 |
 i \hat{R}
 |{\bf p}^\pr s_e s_{\overline{e}}\rangle
 \langle {\bf p}^\pr s_e s_{\overline e}| \hat{{\cal V}}^{^{(0)}}|{\bf p} s_1 
s_2 \rangle\right.
 \nonumber\\
&&~~~~~ \left.-\langle {\bf p}^\pr s_3 s_4 |
 \hat{{\cal V}}^{^{(0)}}
 |{\bf p} s_e s_{\overline{e}}\rangle
 \langle {\bf p} s_e s_{\overline e}|i \hat{R} |{\bf p} s_1 s_2 
\rangle\right)\;.
 \eea
 Recalling Eq.~(\ref{eq:rzero}) and using Eq.~(\ref{eq:hu}) we have
 \bea
 \hard&=&\frac{\alpha}{\pi^2}\int d^3p d^3 p^\pr 
 \langle \phi_{100}|{\bf p}^\pr\rangle\langle{\bf p}|\phi_{100}\rangle
 \frac{F}{\pp}
 \;,
 \eea
 where
 \bea
 F&=&\sum_{s_e s_{\overline{e}}}\langle {\bf p} s_e s_{\overline{e}} |
 i \hat{R}
 |{\bf p} s_1 s_2\rangle
 \langle {\bf p}^\pr s_3 s_4| \hat{v}^{(0)}|{\bf p} s_e s_{\overline{e}} 
\rangle\\
 &=&
 \frac{1}{2} \sum_{s_e s_{\overline{e}}}\left(
 \langle {\bf p} s_e s_{\overline{e}} |
 i \hat{R}
 |{\bf p} s_1 s_2\rangle-\langle {\bf p}^\pr s_e s_{\overline{e}} |
 i \hat{R}
 |{\bf p}^\pr s_1 s_2\rangle
 \right)\langle {\bf p}^\pr s_3 s_4| \hat{v}^{(0)}|{\bf p} s_e s_{\overline{e}} 
\rangle
 \;,
 \eea
 using the fact that $\hat{v}^{(0)}$ is odd under ${\bf p} \longleftrightarrow 
{\bf p}^\pr$ 
 in this last step.
 Using Eq.~(\ref{eq:hu}) this becomes
 \bea
 F&=&\frac{1}{4m}\sum_{s_e s_{\overline{e}}}{ v}^{(0)}({\bf p}^\pr s_3 s_4;
 {\bf p} s_e s_{\overline{e}}){ v}^{(0)}({\bf p}^\pr s_e s_{\overline{e}};
 {\bf p} s_1 s_2)
 \;.
 \eea
 Using the even symmetry of the rest of the integrand under the operations
 $(p_x \longrightarrow - p_x,p_x^\pr \longrightarrow - p_x^\pr)$ and 
 $(p_x \longleftrightarrow p_y, p_x^\pr \longleftrightarrow p_y^\pr)$ this sum
 can be simplified with result
 \bea
 F&=&-\frac{1}{24 m} \left(3 g_1+g_2\right) \pp
 \;,
 \eea
 where
 \bea
g_1&=& s_1 s_3+s_2 s_4~,
\label{eq:g1}\\
g_2&=& 1+s_1 s_2-s_2 s_3-s_1 s_4+s_3 s_4+s_1 s_2 s_3 s_4
\;.\label{eq:g2}
\eea
Recall that $s_i = \pm 1$, $(i=1,2,3,4)$;  the `$\frac{1}{2}$' has been factored 
out of these 
spins.\footnote{
In order to get these simple forms for $g_1$ and $g_2$ 
it was useful to note the following simple relation: 
$\del_{s s^\pr} = \frac{1}{2} s (s+s^\pr)$ 
 (true because $s^2=1$).}
 The result was written in this form to show the equivalence with \cite{jones}.
 Combining the results we have
 \bea
 \hard&=&
 -\frac{\alpha}{24 \pi^2 m} \left(3 g_1+g_2\right) \int d^3 p d^3 p^\pr \langle 
\phi_{100}|{\bf
 p}^\pr\rangle\langle{\bf
 p}|\phi_{100}\rangle \\
 &=&-\frac{m^2 \alpha^4}{24}\left(3 g_1+g_2\right)\;,
 \eea
 using Eq.~(\ref{eq:hy}) in this last step.
 
 Combining the results we have
 \bea
 \frac{\easy+\hard}{2 m^2 \alpha^4}&=& \frac{1}{2} \del_{s_1 s_2} \del_{s_1 s_4} 
\del_{s_3 s_4} 
 -\frac{1}{12} \del_{s_1 {\overline s}_2} \del_{s_3 {\overline s}_4}+\frac{1}{2} 
 \del_{s_2 {\overline s}_4} \del_{s_1 {\overline s}_2} \del_{s_1
  {\overline s}_3}\nonumber\\
  &&~~~~~-\frac{1}{48} (3 g_1+g_2)\;.
 \eea
 The eigenvalues are
  \bea
 \left\langle 1 \left|\easy+\hard\right| 1 \right\rangle&=&- \frac{5}{3} m^2
 \alpha^4\label{eq:shit}~,\\
 \left\langle 2 \left|\easy+\hard\right| 2 \right\rangle&=& \frac{2}{3} m^2 
\alpha^4~,\\
  \left\langle 3 \left|\easy+\hard\right| 3 \right\rangle&=&  \frac{2}{3} m^2 
\alpha^4~,\\
 \left\langle 4 \left|\easy+\hard\right| 4 \right\rangle&=&  \frac{2}{3} m^2 
\alpha^4
 \;,\label{eq:shitt}
 \eea
 with corresponding eigenvectors 
 \bea
&&\left\{ |1\rangle=\frac{|+-\rangle-|-+\rangle}{\sqrt{2}}\;,\;
|2\rangle=\frac{|+-\rangle+|-+\rangle}{\sqrt{2}}\;,\;
|3\rangle=|--\rangle\;,\;
|4\rangle=|++\rangle\right\}\nonumber\:.
 \eea
 
 These results translate to the well known answer, $\frac{7}{6} \alpha^2 Ryd$, 
as can be
 seen by recalling the definition of our zeroth order and exact mass squared:
 \bea
 (2m+B_N)^2&=& 4 m^2+4m{\cal B}_N+ \del^{^{(1)}}M^2+{\cal O}\left(m^2 
\alpha^5\right)\;,
 \eea
 which gives
 \bea
 B_{triplet}-B_{singlet}&=&\frac{7}{6} \alpha^2 Ryd+{\cal O}\left(m 
\alpha^5\right)
 \;,
 \eea
 the desired result.
\newpage
\centerline{{\bf  ACKNOWLEDGMENTS}}
\vskip.2in
The author wishes to thank the organizers of
{\em Orbis Scientiae 1997} for providing such a stimulating atmosphere in which 
to work.
The author would also like to thank  Brent H. Allen, Martina M. Brisudov\'{a},
Stanis{\l}aw D. 
 G{\l}azek, Robert J. Perry, David G. Robertson
  and Kenneth G. Wilson for all the useful discussions that led to some
simplifications in the calculations, and  to a whole lot of understanding.
 In particular,
the author would like to thank Brent H. Allen for the use of his
matrix element rules \cite{brent}. 
Research reported in this paper has been supported  by 
the National Science Foundation under grant PHY--9409042.

\end{document}